\documentclass[showpacs,apl,twocolumn,floatfix]{revtex4}

\usepackage{graphicx,float,amsmath}

\begin{document}

\title{Absence of an intrinsic value for the surface recombination velocity in doped semiconductors}

\author{F. Cadiz}
\affiliation{Physique de la mati\`ere condens\'ee, Ecole Polytechnique, CNRS, 91128 Palaiseau, France}

\author{D. Paget}
\email{daniel.paget@polytechnique.edu}
\affiliation{Physique de la mati\`ere condens\'ee, Ecole Polytechnique, CNRS, 91128 Palaiseau, France}

\author{V. L. Berkovits, V. P. Ulin}
\affiliation{A. F. Ioffe Physico-technical Institute, 194021 Saint Petersburg, Russia}

\author{S. Arscott, E. Peytavit}
\affiliation{Institut d'Electronique, de Micro\'electronique et de Nanotechnologie (IEMN), University of Lille, CNRS, Avenue Poincar\'e, Cit\'e Scientifique, 59652 Villeneuve d'Ascq, France}

\author{A.C.H. Rowe}
\affiliation{Physique de la mati\`ere condens\'ee, Ecole Polytechnique, CNRS, 91128 Palaiseau, France}

\begin{abstract}
A self-consistent expression for the surface recombination velocity $S$ and the surface Fermi level unpinning energy as a function of light excitation power ($P$) is presented for n- and p-type semiconductors doped above the 10$^{16}$ cm$^{-3}$ range. Measurements of $S$ on p-type GaAs films using a novel polarized microluminescence technique are used to illustrate two limiting cases of the model. For a naturally oxidized surface $S$ is described by a power law in $P$ whereas for a passivated surface $S^{-1}$ varies logarithmically with $P$. Furthermore, the variation in $S$ with surface state density and bulk doping level is found to be the result of Fermi level unpinning rather than a change in the intrinsic surface recombination velocity. It is concluded that $S$ depends on $P$ throughout the experimentally accessible range of excitation powers and therefore that no instrinsic value can be determined. Previously reported values of $S$ on a range of semiconducting materials are thus only valid for a specific excitation power. 
\end{abstract}
\maketitle

\section{Introduction}
\label{intro}

Knowledge of surface recombination in semiconductors is crucial for bipolar nanoelectronics \cite{hasegawa2010}, in particular for high surface-to-volume ratio structures such as nanowires \cite{chang2012}. As a result surface recombination has been extensively explored using  photoconductivity \cite{polignano1999,ogita1996}, surface photovoltage measurements \cite{storr1985}, scanning electron microscopy \cite{jastrzebski1977}, cathodoluminescence \cite{wittry1966}, or photoluminescence \cite{mettler1977,ito1994,ahrenkiel2009} on a wide range of materials including GaAs \cite{ito1994}, Si \cite{aberle1992}, InP \cite{moison1986}, GaN \cite{aleksiejunas2003}, InN \cite{cusco2009}, ZnSe \cite{wang1998} and alloys \cite{cui2002}. The immense majority of these works rely on the fact that the surface recombination velocity $S$ is a fundamental parameter for describing the surface electronic properties \cite{aspnes1983}.  To the best of our knowledge only two works, one at the Si/SiO$_2$ interface \cite{aberle1992} and the other on GaAs surfaces \cite{hasegawa2010}, have investigated the excitation power dependence of $S$. From a theoretical point of view, the most comprehensive treatments of the problem rely heavily on computer simulations \cite{hasegawa2010,aberle1992,saitoh1992}. In short, a simple description of the power dependence of $S$ is still lacking.

Here we obtain a simple analytical expression for $S$ as a function of light excitation power for p- or n-type doped semiconductors (above the 10$^{16}$ cm$^{-3}$ range) and we show that $S$ depends strongly on the excitation power, even at very low intensities. From an experimental point of view, $S$ is estimated using an original technique based on imaging of the luminescence under tightly-focussed light excitation \cite{favorskiy2010}.  The measured power dependence of $S$ is in excellent agreement with the predictions of the model and it is concluded that $S$ cannot be viewed as an absolute characterization of surface electronic properties. The paper is organized as follows. In Sec. \ref{theory}, we obtain expressions for $S$ as a function of light excitation power. The experimental method is explained in Sec. \ref{experiment} and the results are presented in Sec. \ref{results} and discussed in Sec. \ref{discussion}.

\section{Theory}
\label{theory}

\begin{figure}[tbp]
\includegraphics[clip,width=8.5 cm] {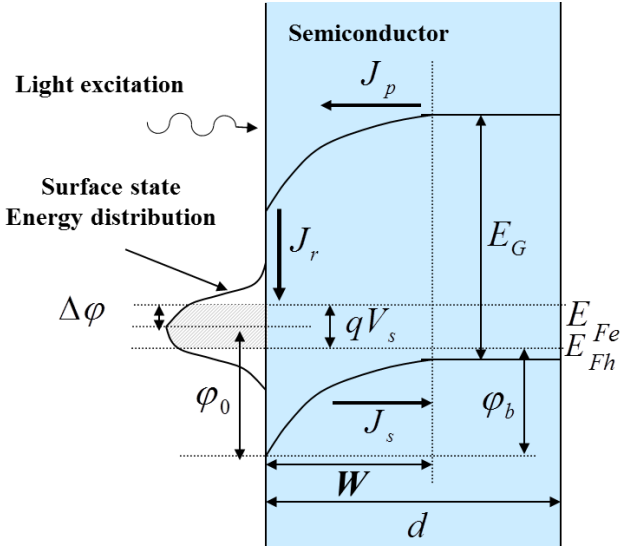}
\caption{A schematic representation of the band structure of the photo-excited GaAs film of thickness $d$ showing the light-induced shift of the surface electron Fermi level, $\Delta\varphi$, and the photovoltage $qV_s$. Also shown are the photocurrent $J_p$, the Schottky current $J_s$ and the recombination current $J_r$.}
\label{figbands}
\end{figure}

This section is devoted to the calculation of $S$. As with previous work \cite{vu2010}, it is based on current conservation and charge neutrality. Fig. \ref{figbands} summarizes the near surface band structure scheme which will be used in the specific case of p-type material. When photo-excited, a steady-state photo-electron concentration is established whose value at the edge of the depletion zone is $n_0$. This results in a photo-current of magnitude $J_p = qn_0S$, where $q$ is the absolute value of the electron charge. Here, the photo-current generated within the depletion zone is neglected, which, for an excitation energy slightly above bandgap, is valid for doping levels larger than several  10$^{16}$ cm$^{-3}$. Injection of photo-electrons creates a surface quasi equilibrium characterized by an electron quasi-Fermi level at energy $qV_s$ above its position in the bulk, where $qV_s$ is the photovoltage shifted by $\Delta\varphi$ with respect to its position in the dark. As seen from Fig. \ref{figbands}, the surface barrier is given by \begin{equation} \label{barrier} \varphi_b = \varphi_0 + \Delta\varphi-qV_s \end{equation} where $\varphi_0$ is the barrier value in the dark and $\Delta\varphi$ and $qV_s$ are related by the equality of $J_p$ with the Schottky current $J_s$ : \begin{equation} \label{currents} qn_0S = J_0\exp\left[-\frac{\Delta\varphi}{k_BT}\right]\left(\exp\left[\frac{qV_s}{k_BT}\right]-1\right). \end{equation} Here the saturation current $J_0$ is related to the effective Richardson constant $A^{**}$ and to the surface barrier in the dark $\varphi_0$ by $J_0 = A^{**}T^2\exp\left[-\varphi_0/k_BT\right]$.

In order to calculate $S$, one assumes that the centers which cause surface recombination are also those responsible for Fermi level pinning. This excludes materials such as InP \cite{moison1986}. One writes that the surface recombination current $J_r$, given by the Stevenson Keyes expression \cite{aspnes1983}, is equal to the photocurrent \begin{align} \label{stephkayes} J_r &= q\int_{E_{Fh}}^{E_{Fe}} N_T(\epsilon) \frac{\sigma_n\nu_n\sigma_p\nu_pn_sp_s}{\sigma_p\nu_p(p_s+p_{ts})+\sigma_n\nu_n(n_s+n_{ts})} \mathrm{d}\epsilon \notag \\ &= qn_0S \end{align} where $N_T(\epsilon)$ is the surface concentration of defects per unit energy at an energy $\epsilon$ above the position of the Fermi level at equilibrium. Here, $\sigma_n(\sigma_p)$ and $\nu_n(\nu_p)$ are the electron (hole) capture cross sections at the centers and thermal velocities, respectively, $n_s(p_s)$ are the volume concentrations of electrons (holes) at the surface and $n_{ts}(p_{ts})$ are their values at equilibrium if the Fermi level coincides with the energy ($\epsilon$) of the surface centers. $J_r$ is obtained by integration over centers situated between the hole and electron quasi Fermi levels since centers situated below $E_{Fh}$ are occupied and centers situated above $E_{Fe}$ are empty. Eq. \ref{currents} implies that surface recombination strongly depends on the photoelectron concentrations at the surface $n_s$ and at the edge of the depletion region $n_0$, respectively. Both $n_s$ and $n_0$ are in turn dependent on the excitation power and/or on surface recombination itself so that Eq. \ref{currents} should be solved self-consistently and cannot be viewed as a fundamental expression for $S$.

Provided the surface barrier is not too small, among the four terms of the denominator of Eq. \ref{stephkayes}, $\sigma_n\nu_nn_s$ is by far the dominant one, so that surface recombination is limited by hole capture. This is justified more precisely in Appendix \ref{appendix} and also implies that the occupation probability of centers between $E_{Fh}$ and $E_{Fe}$ is close to unity. Thus the centers which dominate the surface recombination are those situated at the electron quasi Fermi level, in a typical energy width of $k_BT$. Their volume concentration is $N^*_T(\Delta\varphi) = N_T(\Delta\varphi)k_BT/a$ where $a$ is a typical thickness of the surface layer. Further expressing $n_sp_s = n_i^2\left(\exp\left[qV_s/k_BT\right]-1\right)$ where $n_i$ is the intrinsic concentration, Eq. \ref{stephkayes} becomes \begin{equation} \label{currents2} qn_0S = J_{r0}\frac{N^*_T(\Delta\varphi)}{n_s}\left(\exp\left[\frac{qV_s}{k_BT}\right]-1\right) \end{equation} where the current $J_{r0}$ is given by \begin{equation} \label{currentrecomb} J_{r0} = q\nu_pn_i(an_i\sigma_p). \end{equation} Assuming equilibrium between the surface and the bulk, one has $n_s = n_0\exp[\varphi_b/k_BT]$. Taking for simplicity $N^*_T(\Delta\varphi) = N^*_T(0)$ which is valid at low excitation power if $\Delta\varphi$ is not too large, Eqs. \ref{barrier}, \ref{currents} and \ref{currents2} give \begin{equation} \label{Sneweq} S = S_0\exp\left[-\Delta\varphi/k_BT\right]\left(1-\exp\left[-qV_s/k_BT\right]\right), \end{equation} where \begin{equation} \label{szero} S_0 = \frac{J_0^2}{qN^*_T(0)J_{r0}}\exp\left[\frac{\varphi_0}{k_BT}\right] \end{equation} is the intrinsic surface recombination velocity obtained for negligible Fermi level unpinning ($\Delta\varphi \neq 0$) provided $\exp\left[-qV_s/k_BT\right] \ll 1$.
	 							
The charge neutrality equation is used to express $qV_s$ as a function of $\Delta\varphi$ so that $S$ will be solely expressed as a function of $\Delta\varphi$, the fundamental quantity of the system. The negative light-induced surface charge must compensate the positive excess charge in the depletion layer so that one obtains \begin{equation} \label{charge} W_0N_A\left[\sqrt{\frac{\varphi_b}{\varphi_0}+\frac{n_s}{N_A}\frac{k_BT}{\varphi_0}}-1\right] = N_T(0)\Delta\varphi \end{equation} Here $W_0$ is the equilibrium width of the depletion layer. The second term in the radical of Eq. \ref{charge} is obtained by spatial integration of the photoelectron charge in the depletion layer \cite{kronik1999}.  The surface concentration is found from Eq. \ref{currents} and Eq. \ref{Sneweq} to be $n_s=A^{**}T^2\exp\left[\Delta\varphi/k_BT\right]/qS_0$. Using Eq. \ref{currents} and Eq. \ref{charge}, we obtain \begin{equation} \label{eta} \exp\left[-\frac{qV_s}{k_BT}\right] = K\exp\left[-(1+\eta^{-1})\frac{\Delta\varphi}{k_BT}\right] \end{equation} where the number \begin{equation} \label{eta2} \eta = \frac{\epsilon}{qW_0N_T(0)} \end{equation} is a measure of the ratio of surface charge to the charge in the depletion zone. Here $\epsilon$ is the semiconductor permittivity. The quantity $K$ is given by \begin{equation} \label{K} K = \exp\left[\frac{\Delta\varphi^2}{4\eta^2\varphi_0k_BT}-\gamma\exp\left[\frac{\Delta\varphi}{k_BT}\right]\right] \end{equation} where \begin{equation} \label{gamma} \gamma = \frac{A^{**}T^2}{qS_0N_A} \end{equation} is equal to unity if the electron charge in the depletion layer is negligible with respect to the fixed acceptor charge (second factor of Eq. \ref{K}) and if $\varphi_b - \varphi_0 \ll \varphi_0$ (first factor). Using Eq. \ref{Sneweq} and Eq. \ref{eta} one obtains \begin{equation} \label{S} S = S_0\exp\left[-\frac{\Delta\varphi}{k_BT}\right]\{1-K\exp\left[-(1+\eta^{-1})\frac{\Delta\varphi}{k_BT}\right]\}. \end{equation} This equation is valid provided surface recombination is not limited by thermal injection of minority carriers, otherwise one has $S=v_n \approx 3\times10^7$ cm/s for p-type GaAs.  Again using $J_p = J_r$ and Eq. \ref{S} one also finds, \begin{equation} \label{fin} K\exp\left[-(1+\eta^{-1})\frac{\Delta\varphi}{k_BT}\right] = \left(qn_0S_0/J_0\right)^{-1}. \end{equation} Eqs. \ref{S} and \ref{fin} are used to calculate $\Delta\varphi$ and therefore $S$ as a function of the excess concentration $n_0$. This type of analytical treatment is analogous to the numerical approach used elsewhere \cite{aberle1992}. In itself however, it is insufficient since $n_0$ also depends on $S$.
 
In order to obtain an explicit expression for $S$ on excitation power, $n_0$ must be calculated by solving (in a unipolar regime at low excitation power \cite{paget2012}) the diffusion equation for the photo-electron concentration $n$ in the bulk of the semiconductor \begin{equation} \label{diffusion} g\tau\phi(r)\alpha\exp\left[-\alpha z\right] - n + L^2 \triangle n = 0 \end{equation} where $r$ is the distance to the excitation spot and $z$ is the depth coordinate. Here $L = \sqrt{D\tau}$, related to the diffusion constant $D$ and to the bulk minority carrier lifetime $\tau$, is the diffusion length, $\triangle$ is the Laplacian operator, $\alpha$ is the light absorption coefficient and $g$ is the rate of electron-hole pair creation. The function $\phi(r)$ describes the light profile and is assumed in this model to be constant. Using the boundary conditions $D\partial n/\partial z = S(n)$ for the front surface, and $D\partial n/\partial z = -S'n$ at the back surface of the sample of thickness $d$, it is found that \begin{equation} \label{nzero} n_0 = \beta N_0 = \frac{v_d}{v_d+S}N_0 \end{equation} where $v_d$ is the diffusion velocity which gives a measure of bulk recombination and $N_0$ is an effective photoelectron concentration, independent on both $S$ and $S'$. Assuming for simplicity that $S'$= 0, one has \begin{equation} \label{bignzero} N_0 = \frac{g\alpha\tau}{(\alpha L)^2 - 1}\frac{\alpha L \{\cosh (d/L) - \exp\left[-\alpha d\right]\} -\sinh (d/L)}{\sinh (d/L)} \end{equation} and \begin{equation} \label{thermal} v_d = \frac{D}{L}\tanh (d/L). \end{equation} Because $v_d$ depends on $d$, it is very large for semi-infinite samples and $n_0$ is independent of $S$. As shown below, this is not the case for thin films. Eqs. \ref{S}, \ref{fin} and \ref{nzero} give the following equation \begin{align} \label{bigequation} K\xi\exp\left[\frac{\Delta\varphi}{k_BT}\right]&- \exp\left[\frac{\Delta\varphi}{k_BT}(1+\eta^{-1})\right] - \nonumber \\ &K = -\frac{v_d}{S_0}\exp\left[\frac{\Delta\varphi}{k_BT}(2+\eta^{-1})\right]. \end{align} This equation allows us to calculate $\Delta\varphi$ as a function of the reduced power \begin{equation} \label{xi} \xi = qv_dN_0/J_0. \end{equation} Eq. \ref{bigequation}, together with Eq. \ref{S}, is the fundamental result of the present work. 

It will be seen below that the right hand side of Eq. \ref{bigequation} is negligible because $v_d$ is small with respect to $S_0$. In this case three power regimes can be identified where $S$ has an analytical expression. At very low excitation power ($\xi \ll 1$) one has $\Delta\varphi \ll k_BT$ so that $K \approx 1$. To first order in $\Delta\varphi/k_BT$, we obtain \begin{equation} \label{simpleS} S \approx S_0\xi \end{equation} so that $S$ increases linearly with $N_0$ as observed for silicon \cite{stephens1994}, essentially because of the increase of $(\exp\left[qV_s/k_BT\right]-1)$ in Eq. \ref{currents2} . Another regime occurs at higher excitation power such that $\xi \gg 1$ provided $K$ remains close to unity (i.e. if $\gamma \ll 1$). In this case the last term on the left hand side of Eq. \ref{bigequation} is negligible and one finds \begin{equation} \label{simpleS2} S \approx S_0\xi^{-\eta} \end{equation} so that $S$ \textit{decreases} with $N_0$ as a power law of exponent $-\eta$. At even larger excitation powers a large concentration of photo-electrons accumulates at the surface and $n_s/N_A \gg \varphi_0/k_BT$ implying in Eq. \ref{charge} that $\Delta\varphi \gg k_BT$ and $K \ll 1$. Eq. \ref{bigequation} simplifies to $K\exp\left[-\Delta\varphi/\eta k_BT\right] = 1/\xi$ for which the dominant factor on the left hand side is the double exponential appearing in Eq. \ref{K}. The approximate value of $S$ is then \begin{equation} \label{supersimpleS} S \approx \frac{\gamma_0 v_d}{\ln\xi}, \end{equation} where $\gamma_0$ is given by \begin{equation} \label{gamma0} \gamma_0 = \frac{A^{**}T^2}{qv_dN_A}. \end{equation} The quantity $\gamma_0$ is related to $\gamma$ (Eq. \ref{gamma}) but does not depend on $S_0$ so that in this regime $S$ no longer depends on the surface state density.

\section{Experiment}
\label{experiment}

\subsection{Measurement of $S$}
\label{measureS}

\begin{figure}[tbp]
\includegraphics[clip,width=8.5 cm] {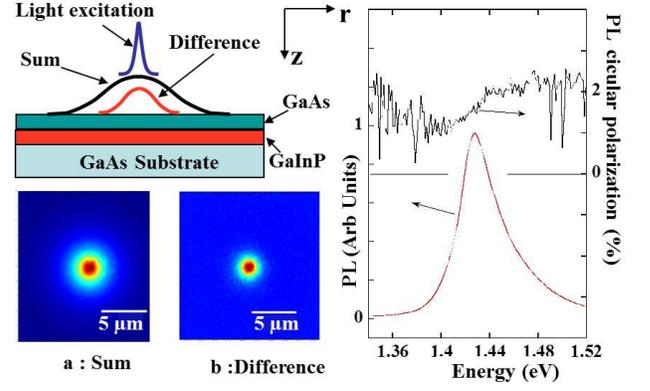}
\caption{Principle of the experiment. The circularly-polarized laser is tightly-focussed on a GaAs film and the spatial distribution of the luminescence intensity and polarization are imaged. Panels a and b show typical sum and difference images, defined by Eq. \ref{imagesum} and Eq. \ref{imagediff}, for oxidized GaAs. Also shown in the right panel are spectra of the photoluminescence and of its degree of circular polarization for the naturally-oxidized surface.}
\label{figexp}
\end{figure}

P-type ($N_A \approx 10^{17}$ cm$^{-3}$) GaAs films of thickness $d = 3\mu$m are photo-excited with circularly-polarized light of wavelength 785 nm. As shown in the left panel of Fig. \ref{figexp}, these films are grown on a GaAs semi-insulating substrate, with a thin GaInP back interface serving as a confinement layer for the photo-electrons and ensuring $S' = 0$ at the back surface.

In order to estimate $S$, a previously described\cite{favorskiy2010} polarized microluminescence technique illustrated in the left panel of Fig. \ref{figexp} is employed. The circularly-polarized light excitation is focussed to a gaussian spot of half width of 0.6 $\mu$m and the resulting luminescence emission and its polarization are spectroscopically analyzed (see right panel of Fig. \ref{figexp}) and imaged (see bottom left panels of Fig. \ref{figexp}). Both $\sigma^+$- and $\sigma^-$- polarized light excitations are in turn used to excite the sample and an image is taken of the $\sigma^\pm$ polarized components of the photo-luminescence with the laser being removed by an appropriate filter. The resulting four images, denoted  $\sigma^{++}$, $\sigma^{+-}$, $\sigma^{--}$ and $\sigma^{-+}$, are combined to form a sum image $I_s = (\sigma^{++} + \sigma^{+-} +\sigma^{--} + \sigma^{-+})/2$ proportional to $n$  and a difference image $I_d = (\sigma^{++} - \sigma^{+-} +\sigma^{--} - \sigma^{-+})/2$ proportional to $n_+ - n_-$. Here, $n_+$ and $n_-$ are the concentrations of electrons with spin aligned parallel or anti-parallel with the direction of light excitation $z$ chosen as the direction of quantization. Typical experimental images for a naturally oxidized GaAs film are shown in panels a and b of Fig. \ref{figexp}, respectively. Their cross sections as a function of radial distance $r$ to the excitation spot are given by \begin{equation} \label{imagesum} I_s(r) = A\int_0^d n(r,z)\exp\left[-\alpha_lz\right] \mathrm{d}z \end{equation} and \begin{equation} \label{imagediff} I_d(r) = A|P_i|\int_0^d (n_+ - n_-)(r,z)\exp\left[-\alpha_lz\right] \mathrm{d}z. \end{equation}

Here, $A$ is a constant and $\alpha_l \approx (3 \mu$m$)^{-1}$ is the absorption coefficient at the luminescence energy. The quantity $P_i$ depends on the matrix elements for recombination and is equal to $\pm0.5$ for $\sigma^{\mp}$ light excitation. The quantity $n_+ - n_-$ is given by the spin diffusion equation \begin{equation} \label{spindiff} (g_+ - g_-)\tau_s\phi(r)\alpha\exp\left[-\alpha z\right] - (n_+ - n_-) + L_s^2\triangle(n_+ - n_-) = 0 \end{equation}
where $(g_+ - g_-)/g = P_i$ and $L_s = \sqrt{D\tau_s}$ is the spin diffusion length. Here $\tau_s$, given by $\tau_s = (1/\tau+1/T_1)^{-1}$ where $T_1$ is the spin-lattice relaxation time, is the spin lifetime. The boundary conditions are the same as for $n = n_+ + n_-$ in Eq. \ref{diffusion}.
 
In the case of charge transport, although Eq. \ref{diffusion} does not have an analytical solution, numerical calculations indicate that for $r > d$, $n$ can be written as $n(r,z) \approx f_1(r)f_2(z)$ where $f_1$ and $f_2$ are independent functions. Standard mathematical treatment then shows that \cite{visvanathan1954} \begin{equation} \label{function} f_1(r) \approx K_0(r/L_{eff}) \approx (r/L_{eff})^{-1/2} \exp\left[-\frac{r}{L_{eff}}\right] \end{equation} where $K_0(u)$ is the modified Bessel function of the second kind, and that $f_2(z)$  has a sinusoidal dependence on $z$ determined by the boundary conditions. One finally obtains \begin{equation} \label{Sdiff} S = D \sqrt{\frac{1}{L_{eff}^2}-\frac{1}{L^2}}\tan\left[\sqrt{\frac{d^2}{L_{eff}^2}-\frac{d^2}{L^2}}\right]. \end{equation}

Because one generally has $D/L_{eff} \ll v_n$, the limit $S \approx v_n$ of very large $S$ is obtained for $L_{eff} \approx 2d/\pi \ll L$ so that the argument of the tangent in Eq. \ref{Sdiff} is $\approx \pi/2$. In order to adapt Eq. \ref{bigequation} for the calculation of $\Delta\varphi$ at distance $r$, it is natural to replace   $N_0$ by $N_0f_1(r)$ in Eq. \ref{xi}, where the value of $f_1(r)$, defined in Eq. \ref{function}, is $I_s(r)/I_s(0)$. One might expect that the strong radial dependence of the electron concentration causes a radial increase of $S$. However, it can be shown that this increase is averaged out. Indeed the spatial dependence of $\Delta\varphi$ induced by spatial variations $N_0f_1(r)$ creates a drift current parallel to the surface in the depletion layer of magnitude $n_s\mu_nW_0\delta\varphi_b/\delta r = qW_0n_sD/L_{eff}$, where $\mu_n = D k_BT/q$ is the electron mobility. This current produces a spatial averaging of $n_s$ and therefore of $S$. The ratio of this current to the diffusion current in the film, $qdN_0f_1(r)D/L_{eff}$, is equal to $(W_0/d)\exp\left[\varphi_b/k_BT\right]$ and is very large. The slope of the experimental curves thus gives a well-defined value of $S$, related to a spatially averaged concentration $N_0\langle f_1(r) \rangle$. This averaging does not perturb i) the investigation of the relative change of surface recombination induced by passivation treatments, provided $S$ is measured at constant values of photoelectron concentrations, ii) the dependence  of $S$ on excitation power, and iii) the value of the exponent $\eta$. 

\subsection{Samples}
\label{samples}

For the samples described in Fig. \ref{figexp}, the bulk charge and spin transport parameters were obtained using a GaInP passivated top surface that originally covered the sample and ensured a negligible surface recombination velocity. Investigations performed before the chemical removal of this surface GaInP layer yielded $L \approx 22 \mu$m and $L_s \approx 1.2 \mu$m \cite{favorskiy2010}. Independent measurement of $\tau$ gives $D$ = 150 cm$^2$/s \cite{paget2012}, from which $T_1 \approx 0.96 \times 10^{-10}$ s is estimated. This value is only a factor of two larger than the value found at 300 K for the same acceptor concentration in bulk GaAs of very distinct origin \cite{zerrouati1988}.

In order to illustrate the limiting cases underlined in Sec. \ref{theory}, five different samples were used, all characterized by the structure shown in Fig. \ref{figexp}, and by the same properties for bulk transport and recombination. The features of these samples and the results are summarized in Table \ref{table1}. Sample A has a naturally oxidized surface and the right panel of Fig. \ref{figexp} shows the photo-luminescence spectrum and its degree of circular polarization for this case. One finds a degree of circular polarization of about 2 \%, thus revealing a photo-electron spin polarization of 4 \%. Sample B was obtained by treating the oxidized surface for 1.5 minutes by a saturated sodium sulfide solution. This treatment is known to saturate Ga surface dangling bonds by sulfur atoms, and to reduce $S$ by about one order of magnitude \cite{paget1996}. In order to illustrate the case of a negligibly small $S$ (sample C), the results previously obtained for the GaInP encapsulated surface, reported in Ref. \onlinecite{favorskiy2010}, are used. The case of infinite $S$ is illustrated by a surface (sample D) obtained after extensive HCl pre-treatment, known to induce significant degradation of the surface \cite{kang1997}.  Finally, sample E is identical to the GaInP-encapsulated sample C, but was investigated 3 years after growth in order to reveal ageing effects in the GaInP overlayer. A distinct piece of this sample, labelled E', was used for the investigation of the effect of light excitation power.

\begin{table}[tp]\footnotesize
\caption{Summary of the measured values of the effective diffusion lengths for the samples shown in Fig. \ref{figdiff} at a power of 1 $\mu$W. Estimated values of $S$, as obtained from Eq. \ref{Sdiff}, are also shown.}
\label{table1}\centering 
\begin{tabular}{cccc}

sample & details & $L_{eff}$ ($\mu$m) & $S$ (cm/s) \\ \hline
A & naturally oxidized & 2.2 & $3 \times 10^6$ \\ \hline
B & Na$_2$S passivation & 4.2 & $5.1 \times 10^5$ \\ \hline
C & Encapsulation by 50 nm of GaInP & 21.3 & negligible \\ \hline
D & HCl treatment & 2 & $> 10^7$ \\ \hline
E & Same as C after 3 years & 8.9 & $4.8 \times 10^4$ \\ \hline
\end{tabular}
\end{table}

\section{Results}
\label{results}

\subsection{Effect of passivation}
\label{passivation}

\begin{figure}[tbp]
\includegraphics[clip,width=8.5 cm] {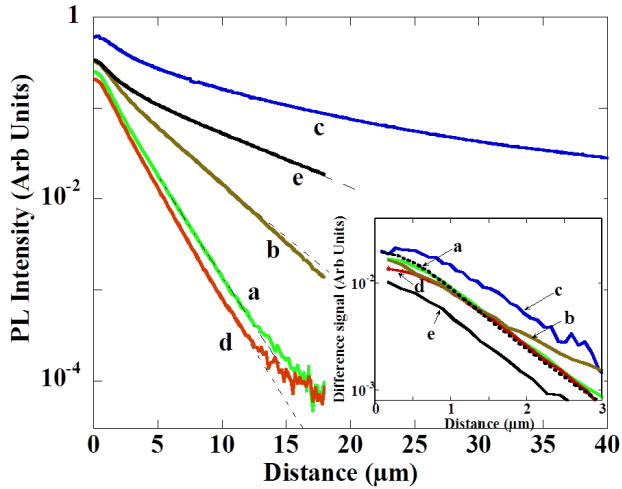}
\caption{Charge diffusion lengths and their fits by Bessel functions in order to determine $L_{eff}$ and $S$ for the naturally oxidized surface (Curve a), and after treatment of the oxidized surface by Na$_2$S (Curve b). The GaInP encapsulated surface (Curve c, taken from Ref. \onlinecite{favorskiy2010} and shifted for clarity), corresponds to $L_{eff}$ = 21.3 $\mu$m and to a very small value of $S$. The surface treated by HCl and hydrazine sulfide (Curve d) corresponds to $L_{eff}$ = 2 $\mu$m and to a very large value of $S$. Finally, the curve for the GaInP encapsulated surface about 3 years after growth (Curve b) reveals slow degradation of the GaInP passivation. As shown in the inset, and unlike the latter sum profiles, the difference profiles are quite similar, because of the dominant effect of spin-lattice relaxation over surface recombination. The numerical calculation of the difference signal using the parameters of the oxidized surface corresponds very well, (dashed line) with no adjustable parameter, to the experimental curve a.}
\label{figdiff}
\end{figure}

The dependence of $S$ on surface treatment is shown in Fig. \ref{figdiff} which presents the angular-averaged cross section of the sum image obtained for different surface terminations of the same GaAs films. In all cases, as seen from the dotted lines, $L_{eff}$ was determined using Eq. \ref{function} in a spatial range corresponding to identical photo-electron concentrations and $S$ was found using Eq. \ref{Sdiff}. In the case of sample A (curve a, taken for a light excitation power of 1 $\mu$W) $S = 3.0 \times 10^6$ cm/$S$ which is close to $4.2 \times 10^6$ cm/$S$ reported in Ref. \onlinecite{ito1994} for an oxidized surface and a similar acceptor concentration. For sample B, curve b reveals a reduction in $S$ of about one order of magnitude with respect to the oxidized surface as expected \cite{paget1996}. The extreme case of negligible $S$ reveals the bulk properties of the GaAs and is shown in curve c for the GaInP-capped surface. This reproduces the results of Ref. \onlinecite{favorskiy2010}. For sample D we find $L_{eff} = 1.9 \mu$m $\approx 2d/\pi$ which, according to Eq. \ref{Sdiff}, corresponds to the case of infinite $S$. This sample illustrates the case of $S \approx v_n$. Curve e corresponds to $L_{eff}= 8.9$ $\mu$m and reveals an increased $S = 4.8 \times 10^4$ cm/s with respect to sample C. This ageing effect has not, to the best of our knowledge, been previously reported for GaInP terminated surfaces, but could be due to the slow evolution of the existing GaInP disorder \cite{leong1996}, which is known to strongly affect the electronic properties \cite{sasaki2001}. 

Shown in the inset of Fig. \ref{figdiff} are the cross sections of the difference images corresponding to curves a-e of the main figure. Immediately apparent is the fact that, unlike charge transport, spin transport depends only weakly on surface recombination and consequently the cross sections of the corresponding difference images are quite similar. This is expected since the effective spin lifetime is limited by spin-lattice relaxation (rather than by surface recombination) which only weakly depends on the surface treatment. For a quantitative interpretation, the analytical treatment summarized in Eqs. \ref{imagesum} and \ref{imagediff} is not valid since the distance to the excitation spot is comparable with the sample thickness. The dashed curve in the inset of Fig. \ref{figdiff} is a numerical solution of the spin diffusion equation for the case of the oxidized surface. This curve, obtained with no adjustable parameters, corresponds very well with the experimental curve a. 

\subsection{Effect of light excitation power}
\label{power}

\begin{figure}[tbp]
\includegraphics[clip,width=8.5 cm] {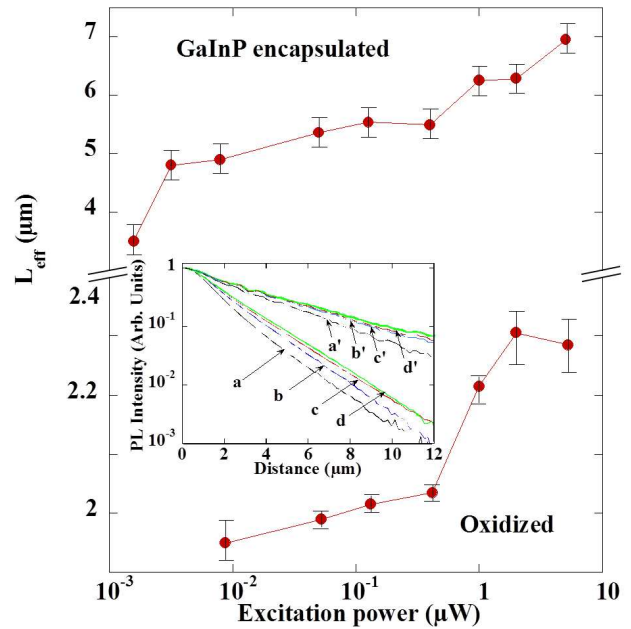}
\caption{Effect of excitation light power on the measured value of $S$. The inset shows the charge distribution profile for the oxidized sample for selected excitation powers : (a) $10^{-2}$ $\mu$W, (b) 0.45 $\mu$W, (c) 1.1 $\mu$W and (d) 5.7 $\mu$W, and for the GaInP encapsulated surface for the same excitation powers (curves a' to d', respectively). The main figure shows the dependences as a function of excitation power of the calculated values of $L_{eff}$.}
\label{figpower}
\end{figure}

\begin{figure}[tbp]
\includegraphics[clip,width=8.5 cm] {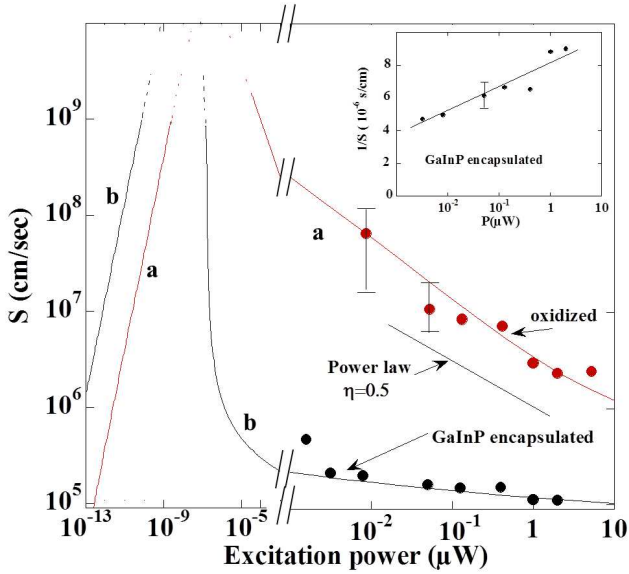}
\caption{The data points show the excitation power dependence of $S$, obtained from the results of Fig. \ref{figpower} using Eq. \ref{Sdiff}. Note that the electronic concentration at which $S$ is measured would correspond to a homogeneous excitation power about two orders of magnitude smaller. The solid lines show dependences calculated using Eq. \ref{S} and Eq. \ref{bigequation}. For the oxidized surface,(Curve a) the dependence is close to a power law, as predicted by Eq. \ref{simpleS2} for $\eta \approx 0.5$, which gives a density of surface states $N_Y \approx 10^{12}$ cm$^{-2}$eV$^{-1}$. For the GaInP-encapsulated surface(Curve b), $S$ is almost independent of $P$, because of the presence of a significant photoelectron concentration in the depletion layer.  Also shown are extremely low (experimentally unreachable) powers, $P < 10^{-9}$ $\mu$W, at which $S$ is predicted to increase with $P$. The inset shows that for the GaInP encapsulated sample, $S^{-1}$ decreases linearly as a function of the logarithm of the power. }
\label{figpowerS}
\end{figure}

The effect of light excitation power has been investigated between an extremely small value of about 1 nW and 5.7 $\mu$W for the samples A and  E'. The inset of Fig. \ref{figpower} shows the sum cross sections for the two samples for selected excitation powers and reveals an increase of $L_{eff}$ with excitation power. This increase, from 1.9 $\mu$m to 2.3 $\mu$m for the oxidized surface, and from 3.5 $\mu$m to 7 $\mu$m for the encapsulated one. The corresponding values of $S$, calculated using Eq. \ref{Sdiff}, are shown in Fig. \ref{figpowerS}.  For the oxidized surface at high power, one has $S = 2 \times 10^6$ cm/s, corresponding to the result of Fig. \ref{figdiff}. A decrease in the power results in an increase in $S$ by more than one order of magnitude so that, at low power, $L_{eff}$ is close to $2d/\pi$ corresponding to the $S \approx v_n$ situation characterized by sample D.  For the GaInP-encapsulated surface on the other hand, $S$ increases only by about a factor of 2 with an equivalent power decrease.

\begin{figure}[tbp]
\includegraphics[clip,width=8.5 cm] {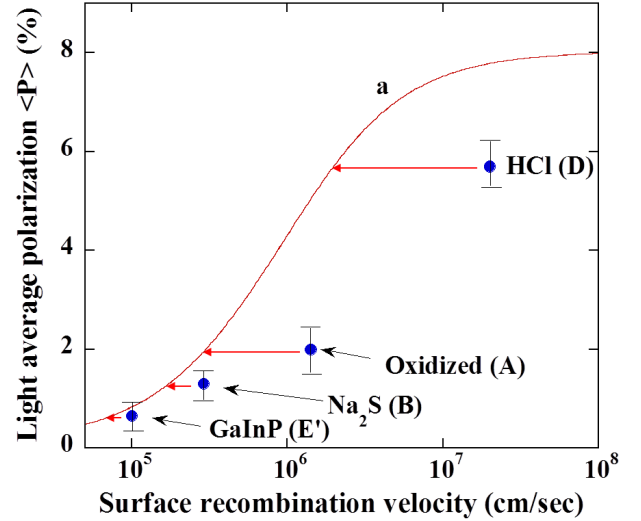}
\caption{Spatially-integrated luminescence degree of circular polarization as a function of $S$ measured for $P = 1$ $\mu$W and at $r \approx 10$ $\mu$m. Curve a is obtained using a numerical resolution of the charge and spin diffusion equations. The experimental results of Fig. \ref{figdiff} and Fig. \ref{figpower} systematically correspond to $S$ higher than the result of the calculation. This demonstrates the reduction of surface recombination velocity at the center, by an amount marked by arrows.}
\label{figlum}
\end{figure}

By monitoring the spatially averaged degree of circular polarization of the luminescence, defined as $\langle P \rangle = \langle I_d(r) \rangle/\langle I_s(r) \rangle$, higher excitation powers can be explored. Indeed $\langle P \rangle$ is dominated by $n_+ - n_-$ and $n$ at $r = 0$, larger than their values at 10 $\mu$m by more than one order of magnitude. Shown in Fig. \ref{figlum} are the dependences of $\langle P \rangle$ for the selected surfaces shown in Fig. \ref{figdiff} and Fig. \ref{figpower} as a function of the values of $S$ that are experimentally determined from the sum cross sections at $r = 10$ $\mu$m. Curve a shows the calculated dependence using a one dimensional resolution of the charge and spin diffusion equations for several values of $S$. For all data points, the surface recombination velocity at $r = 10$ $\mu$m is larger (by up to 1 order of magnitude) than the value deduced from $\langle P \rangle$. For the oxidized surface, $S = 3.0 \times 10^5$ cm/s corresponds well with an extrapolation of curve a of Fig. \ref{figpowerS} to an excitation power of 1 mW. Conversely, for sample E', the $S$ almost coincides with its value calculated at $r = 0$, in agreement with the weak power dependence of $S$ for this sample \cite{paget2012}. As expected, the sulfide-passivated sample exhibits an intermediate behavior. 

\section{Discussion}
\label{discussion}

\subsection{Comparison with the predictions of the model}
\label{model}

The experimental results are in excellent agreement with the model of Sec. \ref{theory}. From a semi-quantitative point of view, the dependence of $S$ as a function of excitation power for the oxidized sample agrees with the power law predicted by Eq. \ref{simpleS2} with an exponent of the order of 0.5. This gives $N_T(0)$ of the order of several 10$^{12}$ cm$^{-2}$/eV which lies in the typical range of experimentally observed surface state concentrations \cite{hasegawa2010}. The smaller power dependence of $S$ for the encapsulated sample is viewed as the very high power regime illustrated by Eq. \ref{supersimpleS}. As shown in the inset of Fig. \ref{figpowerS}, the inverse recombination velocity is approximately proportional to the logarithm of the excitation power. With respect to the oxidized sample, Eq. \ref{simpleS2} would predicts a stronger power dependence because of the reduced value of $N_T(0)$. However, this effect is masked by the increase of the surface photo-electron concentration resulting from a reduction in $S$. As a result $K$ decreases because the second factor in Eq. \ref{K} can be written $\exp\left[\gamma_0v_d/S\right]$ and Eq. \ref{supersimpleS} must be applied.

A more quantitative verification of the model of Sec. \ref{theory} is provided by the solid lines of Fig. \ref{figpowerS}, which are the power dependences of $S$ calculated numerically using Eq. \ref{S} and Eq. \ref{bigequation}. Keeping all parameter values (apart from $N_T$) unchanged for the two samples, the agreement with the experimental results for the two samples is highly satisfactory. We have used $N_T(0) = 10^{12}$ cm$^{-2}$/eV, as suggested above, and $S_0 = 4\times10^{10}$ cm/s for the oxidized surface. Note that although $S_0$ may appear very large, the real value of $S$ is much smaller. For the encapsulated surface (E'), $N_T \approx 3 \times 10^{10}$ cm$^{-2}$/eV is used and $S_0 = 4 \times 10^{10} (N_T^0/N_T)$ cm/s.  For the two samples we have taken $\xi = 10^6$ for a power of 1 $\mu$W at $r = 10$ $\mu$m from the light spot, and $\gamma_0 = 180$. Finally, the quantities $N_0$ and $v_d \approx 10^4$ cm/s were calculated using Eq. \ref{bignzero} and Eq. \ref{thermal} respectively. It is seen that this choice of parameters gives acceptable values for three key quantities. For the surface barrier, expressed as $\varphi_0 = k_BT\ln\left[N_A\gamma_0\xi/N_0\right]$ we find 0.61 eV. The effective Richardson constant, obtained from $A^{**} = qT^{-2}v_dN_A\gamma_0$, is found equal to $3 \times 10^3$ Am$^{-2}$K$^{-2}$ which is a factor of 30 smaller than the value of the unreduced constant $A$ for light holes. The difference can be attributed to the reduction of Schottky current due to the reduction in the probability of a majority carrier to reach the surface \cite{rhoderick1982}. Using Eq. \ref{szero}, the value of $S_0$ is related to the hole capture cross section, expressed as \begin{equation} \label{holecross} \sigma_p = (N_T(0)k_BT)^{-1} \frac{v_d^2}{S_0v_p} \frac{N_AN_0}{n_i^2}\frac{\gamma_0}{\xi}. \end{equation} We estimate $\sigma_p \approx 10^{-6}$ cm$^2$. Although the accuracy is poor (the range of values of $S_0$, $\gamma_0$ and $\xi$ which allow us to interpret the data is of about 1 or 2 orders of magnitudes) this value is significantly larger than values of about 10$^{-15}$ cm$^2$ measured by capacitance transient spectroscopy for bulk defects \cite{henry1977}. However, the present value may not seem unrealistic for at least three reasons: i) it appears that trapping cross sections at defects of oxidized GaAs can be much larger than for bulk defects. Values as high as 10$^{-9}$ cm$^2$ have been reported \cite{deenapanray2003}, ii) these values can further be increased by the probable presence of nanoclusters of elemental As \cite{brunkov2000}, in particular for the surface prepared by HCl decapping of GaInP \cite{kang1997}, and iii) trapping of majority carriers in the flat band conditions used for capacitance spectroscopy overlooks tunnelling-assisted trapping over the top of the barrier. Such an effect will obviously increase the effective concentration $p_s$ of recombining majority carriers and induce a decrease of their energy by an amount $\delta E$. In Eq. \ref{stephkayes}, such an effect can be taken into account by an effective decrease of the bandgap and an increase of $n_i^2$ by $\exp\left[\delta E/k_BT\right]$, thus reducing the effective value of $\sigma_p$. Using Ref. \onlinecite{rhoderick1982}, we calculate that values of $\delta E$ of the order of 30 \% of the barrier are completely realistic, leading to a further decrease of $\sigma_p$ by more than 3 orders of magnitude. 

\begin{figure}[tbp]
\includegraphics[clip,width=8.5 cm] {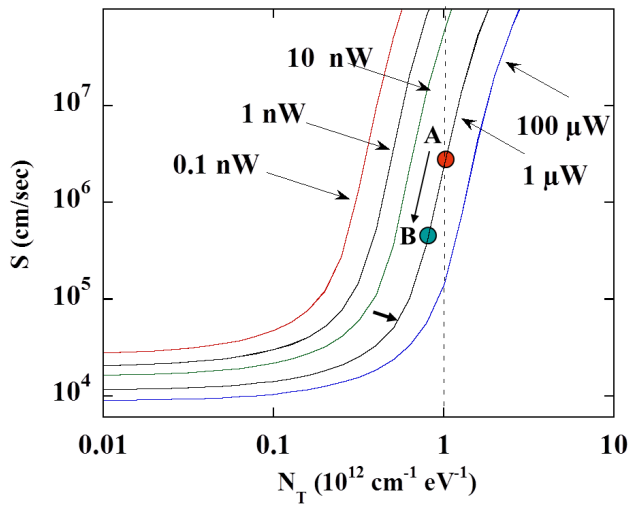}
\caption{Calculated dependence of $S$ as a function of $N_T$ for selected excitation powers. Dot A shows the situation illustrated by curve a of Fig. \ref{figdiff} (oxidized surface). One sees that the order of magnitude decrease of $S$ induced by sulfide passivation (Fig. \ref{figpower}, curve b, dot B in this figure) is caused by only a 25 \% decrease of $N_T$. For realistic values of the surface state density and light excitation power this change is only weakly dependent on power.}
\label{fignt}
\end{figure}

\subsection{Dependence of $S$ on trap concentration and power density}
\label{dependence}

In addition to being able to correctly interpret the experimental results on GaAs films, as shown in Appendix, the model summarized by Eq. \ref{simpleS}, Eq. \ref{simpleS2} and Eq. \ref{supersimpleS} is applicable to a wide class of semiconducting n- or p-type doped materials. As an example, the present model is in agreement with experimental results obtained on p-type silicon, for which a power law is also observed \cite{stephens1994}.  Using the results of the latter work at a resistivity of 10 $\Omega$cm, for which the precision is sufficient, we calculate using Eq. \ref{eta2}, $N_T(0) = 2 \times10^{12}$ cm$^{-2}$eV$^{-1}$, i.e. only a factor of 2 larger than the value obtained using a complete numerical calculation \cite{aberle1992}.

These findings have three main implications for surface recombination studies which are at variance with the general picture of Ref. \onlinecite{aspnes1983} in the particular case of doped semiconductors.  Firstly, the key role of light excitation power in the determination of $S$ has been demonstrated even for extremely low powers. As a result, most values of $S$ reported in the literature using photo-luminescence are likely to be underestimated because of the effect of Fermi level unpinning and cannot be considered as absolute determinations. As an example, the results of Fig. \ref{figpowerS} have been obtained with a value of $N_0f_1(r)$ similar to the one used in Ref. \onlinecite{ito1994}. While the value of $S$ is very close to that obtained by Ref. \onlinecite{ito1994} for the same acceptor concentration, we conclude that a higher values of $S$, limited by the electron thermal velocity, would have been obtained for a reduced power level. If the power is reduced to extremely small values ($P < 10^{-15}$ W for GaAs), $S$ would decrease to become smaller than $v_n$ as shown in Fig. \ref{figpowerS}. While these small powers are experimentally unrealistic for GaAs, this regime has already been observed in silicon \cite{stephens1994}. Thus the quantity $S_0$, which may appear to be the photo-electron concentration independent (and therefore fundamental) value of $S$, can only be obtained from a detailed analysis of the power dependence of $S$.

Secondly, the change of $S$ induced by passivation, in contradiction with generally accepted ideas summarized by Eq. \ref{currents2}, does not directly reflect the change of surface state concentration since $S_0$ \textit{increases} with decreasing $N_T(0)$. As shown in Fig. \ref{figpowerS}, this effect can only be observed at unattainably low excitation powers in the regime described by Eq. \ref{simpleS}. Fig. \ref{fignt} shows the calculated dependence of $S$ as a function of $N_T$ for selected, realistic, excitation powers. Immediately apparent is the fact that, in contrast with the previously accepted ideas, upon passivation, $N_T(0)$ decreases much less than $S$. Indeed, we calculate that the decrease of $S$ by one order of magnitude produced by the sulfide passivation in Fig. \ref{figdiff} implies a decrease of $N_T(0)$ by only 25\%. Further decrease of $N_T(0)$ causes a switching to the regime described by Eq. \ref{supersimpleS} which is shown by the arrow in Fig. \ref{fignt}, below which $S$ only weakly depends on $N_T(0)$. It is concluded that the effect of a change in $N_T(0)$ on $S_0$, given by Eq. \ref{szero}, is masked by that of the change in $\Delta\varphi$.
  
Thirdly, the model presented here is able to qualitatively explain the dependence of $S_0$ on dopant concentration above 10$^{16}$ cm$^{-3}$. For doping levels smaller than about 10$^{16}$ cm$^{-3}$, the approximate proportionality between $S$ and $N_A$ or $N_D$ depending on the semiconductor type \cite{ito1994}, has been interpreted as resulting solely from Eq. \ref{stephkayes} which can be alternatively written for p-type material \begin{equation} \label{stephkayes2} S \propto N^*_T(0)\frac{\sigma_n v_n p_s}{n_0} = N^*_T(0)\frac{\sigma_n v_n N_A}{n_0}\exp\left[-\frac{\varphi_b}{k_BT}\right]. \end{equation} However, for the larger doping concentrations considered by the model, one observes a saturation of this dependence \cite{ito1994}. This result cannot be explained by Eq. \ref{stephkayes2} which would rather give a superlinear dependence because of the increased $\sigma_n$ caused by the localization of minority carriers in the triangular potential well near the surface. The model presented here predicts that the $N_A$ dependence of $S$, as defined by Eq. \ref{simpleS2}, is a combination of the decrease of $\xi^{-\eta}$ ($\eta \propto \sqrt{N_A}$) and of the increase of $S_0$ resulting from the increase of $\sigma_p$ cause by tunnel-assisted trapping processes.

\section{Conclusion}
\label{conclusion}

The theoretical and experimental investigation of surface recombination in doped GaAs films presented here shows that the power dependence of $S$ due to Fermi level unpinning exhibits three phases. At extremely low power levels, $S$ is predicted to increase with excitation power. Although these levels are too low to be reachable for GaAs, this increase has been observed for silicon \cite{aberle1992}. For a naturally oxidized surface at realistic excitation powers $S$ is shown experimentally and theoretically to decrease as a power law for which the exponent depends on surface state density. The GaInP encapsulated surface illustrates a distinct regime where the photo-electron concentration becomes significant with respect to the acceptor concentration because $S$ is reduced. In this case $S^{-1}$ depends logarithmically on power.

The following conclusions, valid for n- or p-type material of doping level larger than 10$^{16}$ cm$^{-3}$ and provided the surface barrier is not too small and that surface recombination is dominated by the states which pin the Fermi level, are drawn. i) No fundamental value of $S$, independent of excitation power, can be directly determined experimentally. This means that for characterizing the surface recombination in devices, $S$ must be measured in exactly the same conditions as under operation of the device. ii) $S$ decreases with increasing surface state density in the intermediate and high power ranges. The actual value of $S$ is not proportional to the surface center concentration, since a relatively marginal passivation of surface centers can lead to a significant decrease of $S$.  

\acknowledgements{Two of us (V. L. B. and V. P. U.) are grateful to the French embassy in Moscow for sponsoring their research stay in France. The authors are grateful to G. Monnier and C. Roberts-Goumet for their participation in some experiments.}

\appendix
\section{Range of validity of the present model}
\label{appendix}

Since we have neglected the photocurrent generated in the depletion layer, the present model is valid, for direct bandgap material, for a doping level larger than about 10$^{16}$ cm$^{-3}$. The present model can be easily extended to n-type material, provided that surface recombination. It suffices to interchange majority and minority carriers and to use $J_{r0} = qv_nn_i(an_i\sigma_n)$. The surface barrier, taken here to 0.6 eV should not be too small, otherwise the surface recombination will no longer be limited by majority carrier trapping. This is shown from the relative magnitudes of the four terms of Eq. \ref{stephkayes}, given by for p-type material, by \begin{equation} \label{app1} \frac{n_{ts}}{n_s} = \frac{qN_cS_0}{A^{**}T^2}\exp\left[-\frac{E_G-\varphi_0}{k_BT} \right]\approx 10^{-9}, \end{equation}

\begin{align} \label{app2} \frac{\sigma_pv_pp_{ts}}{\sigma_nv_nn_s} &= \frac{\sigma_pv_p}{\sigma_nv_n}\frac{qN_vS_0}{A^{**}T^2}\exp\left[-\frac{\varphi_0+2\Delta\varphi}{k_BT}\right] \notag \\ &\approx 7\times 10^{-5} \frac{\sigma_pv_p}{\sigma_nv_n}\exp\left[-\frac{2\Delta\varphi}{k_BT}\right] \end{align} and \begin{align} \label{app3} \frac{\sigma_pv_pp_{s}}{\sigma_nv_nn_s} &= \frac{\sigma_pv_p}{\sigma_nv_n}\frac{qN_AS_0}{A^{**}T^2}\exp\left[-\frac{\varphi_0+2\Delta\varphi}{k_BT}\right] \notag \\ &\approx 8\times 10^{-7} \frac{\sigma_pv_p}{\sigma_nv_n}\exp\left[-\frac{\Delta\varphi-\varphi_0+\varphi_b}{k_BT}\right] \end{align} as a function of  effective densities of states in the valence band $N_v$ and conduction band $N_c$, respectively. The above numerical values are obtained by taking $\sigma_p/\sigma_n \approx 0.01$ \cite{aberle1992}, and the values of $S_0/A^{**}T^2$ and of $\varphi_0$ obtained in Sec. \ref{discussion} for the oxidized surface. As shown above, these ratios are indeed smaller than unity by several orders of magnitude. Conversely, for n-type material, recombination is also limited by trapping of majority carriers since $\sigma_pv_pp_s$ is found to be the dominant term. Both for n- and p-type materials, these ratios strongly increase upon reduction of $\varphi_0$. For p-type material, the ratios of Eq. \ref{app2} and Eq. \ref{app3} are larger than unity provided $\varphi_0 >$ 0.3 eV below which the present model does not apply.

\bibliographystyle{apsrev}
\bibliography{c:/Users/ar/Documents/Publications/References}

\begin{thebibliography}{33}
\expandafter\ifx\csname natexlab\endcsname\relax\def\natexlab#1{#1}\fi
\expandafter\ifx\csname bibnamefont\endcsname\relax
  \def\bibnamefont#1{#1}\fi
\expandafter\ifx\csname bibfnamefont\endcsname\relax
  \def\bibfnamefont#1{#1}\fi
\expandafter\ifx\csname citenamefont\endcsname\relax
  \def\citenamefont#1{#1}\fi
\expandafter\ifx\csname url\endcsname\relax
  \def\url#1{\texttt{#1}}\fi
\expandafter\ifx\csname urlprefix\endcsname\relax\def\urlprefix{URL }\fi
\providecommand{\bibinfo}[2]{#2}
\providecommand{\eprint}[2][]{\url{#2}}

\bibitem[{\citenamefont{Hasegawa et~al.}(2010)\citenamefont{Hasegawa, Akazawa,
  Domanowska, and Adamowicz}}]{hasegawa2010}
\bibinfo{author}{\bibfnamefont{H.}~\bibnamefont{Hasegawa}},
  \bibinfo{author}{\bibfnamefont{M.}~\bibnamefont{Akazawa}},
  \bibinfo{author}{\bibfnamefont{A.}~\bibnamefont{Domanowska}},
  \bibnamefont{and}
  \bibinfo{author}{\bibfnamefont{B.}~\bibnamefont{Adamowicz}},
  \bibinfo{journal}{Applied Surface Science} \textbf{\bibinfo{volume}{256}},
  \bibinfo{pages}{5698} (\bibinfo{year}{2010}).

\bibitem[{\citenamefont{Chang et~al.}(2012)\citenamefont{Chang, Chi, Yao,
  Huang, Chen, Theiss, Bushmaker, LaLumondiere, Yeh, Povinelli
  et~al.}}]{chang2012}
\bibinfo{author}{\bibfnamefont{C.}~\bibnamefont{Chang}},
  \bibinfo{author}{\bibfnamefont{C.}~\bibnamefont{Chi}},
  \bibinfo{author}{\bibfnamefont{M.}~\bibnamefont{Yao}},
  \bibinfo{author}{\bibfnamefont{N.}~\bibnamefont{Huang}},
  \bibinfo{author}{\bibfnamefont{C.}~\bibnamefont{Chen}},
  \bibinfo{author}{\bibfnamefont{J.}~\bibnamefont{Theiss}},
  \bibinfo{author}{\bibfnamefont{A.}~\bibnamefont{Bushmaker}},
  \bibinfo{author}{\bibfnamefont{S.}~\bibnamefont{LaLumondiere}},
  \bibinfo{author}{\bibfnamefont{T.}~\bibnamefont{Yeh}},
  \bibinfo{author}{\bibfnamefont{M.}~\bibnamefont{Povinelli}},
  \bibnamefont{et~al.}, \bibinfo{journal}{Nano Letters}
  \textbf{\bibinfo{volume}{12}}, \bibinfo{pages}{4484} (\bibinfo{year}{2012}).

\bibitem[{\citenamefont{Polignano et~al.}(1999)\citenamefont{Polignano,
  Bellafiore, Caputo, Caricato, Modelli, and Zonca}}]{polignano1999}
\bibinfo{author}{\bibfnamefont{M.}~\bibnamefont{Polignano}},
  \bibinfo{author}{\bibfnamefont{N.}~\bibnamefont{Bellafiore}},
  \bibinfo{author}{\bibfnamefont{D.}~\bibnamefont{Caputo}},
  \bibinfo{author}{\bibfnamefont{A.}~\bibnamefont{Caricato}},
  \bibinfo{author}{\bibfnamefont{A.}~\bibnamefont{Modelli}}, \bibnamefont{and}
  \bibinfo{author}{\bibfnamefont{R.}~\bibnamefont{Zonca}},
  \bibinfo{journal}{Journal of The Electrochemical Society}
  \textbf{\bibinfo{volume}{146}}, \bibinfo{pages}{4640} (\bibinfo{year}{1999}).

\bibitem[{\citenamefont{Ogita}(1996)}]{ogita1996}
\bibinfo{author}{\bibfnamefont{Y.}~\bibnamefont{Ogita}},
  \bibinfo{journal}{Journal of Applied Physics} \textbf{\bibinfo{volume}{79}},
  \bibinfo{pages}{6954} (\bibinfo{year}{1996}).

\bibitem[{\citenamefont{Storr and Haneman}(1985)}]{storr1985}
\bibinfo{author}{\bibfnamefont{G.}~\bibnamefont{Storr}} \bibnamefont{and}
  \bibinfo{author}{\bibfnamefont{D.}~\bibnamefont{Haneman}},
  \bibinfo{journal}{Journal of Applied Physics} \textbf{\bibinfo{volume}{58}},
  \bibinfo{pages}{1677} (\bibinfo{year}{1985}).

\bibitem[{\citenamefont{Jastrzebski et~al.}(1977)\citenamefont{Jastrzebski,
  Gatos, and Lagowski}}]{jastrzebski1977}
\bibinfo{author}{\bibfnamefont{L.}~\bibnamefont{Jastrzebski}},
  \bibinfo{author}{\bibfnamefont{H.}~\bibnamefont{Gatos}}, \bibnamefont{and}
  \bibinfo{author}{\bibfnamefont{J.}~\bibnamefont{Lagowski}},
  \bibinfo{journal}{Journal of Applied Physics} \textbf{\bibinfo{volume}{48}},
  \bibinfo{pages}{1730} (\bibinfo{year}{1977}).

\bibitem[{\citenamefont{Wittry and Kyser}(1966)}]{wittry1966}
\bibinfo{author}{\bibfnamefont{D.}~\bibnamefont{Wittry}} \bibnamefont{and}
  \bibinfo{author}{\bibfnamefont{D.}~\bibnamefont{Kyser}}, in
  \emph{\bibinfo{booktitle}{Proceedings of the International Conference of the
  Physics of Semiconductors, Kyoto}} (\bibinfo{year}{1966}).

\bibitem[{\citenamefont{Mettler}(1977)}]{mettler1977}
\bibinfo{author}{\bibfnamefont{K.}~\bibnamefont{Mettler}},
  \bibinfo{journal}{Applied Physics A: Materials Science \& Processing}
  \textbf{\bibinfo{volume}{12}}, \bibinfo{pages}{75} (\bibinfo{year}{1977}).

\bibitem[{\citenamefont{Ito and Ishibashi}(1994)}]{ito1994}
\bibinfo{author}{\bibfnamefont{H.}~\bibnamefont{Ito}} \bibnamefont{and}
  \bibinfo{author}{\bibfnamefont{T.}~\bibnamefont{Ishibashi}},
  \bibinfo{journal}{Japanese Journal of Applied Physics}
  \textbf{\bibinfo{volume}{33}}, \bibinfo{pages}{88} (\bibinfo{year}{1994}).

\bibitem[{\citenamefont{Ahrenkiel and Johnston}(2009)}]{ahrenkiel2009}
\bibinfo{author}{\bibfnamefont{R.}~\bibnamefont{Ahrenkiel}} \bibnamefont{and}
  \bibinfo{author}{\bibfnamefont{S.}~\bibnamefont{Johnston}},
  \bibinfo{journal}{Solar Energy Materials and Solar Cells}
  \textbf{\bibinfo{volume}{93}}, \bibinfo{pages}{645} (\bibinfo{year}{2009}).

\bibitem[{\citenamefont{Aberle et~al.}(1992)\citenamefont{Aberle, Glunz, and
  Warta}}]{aberle1992}
\bibinfo{author}{\bibfnamefont{A.}~\bibnamefont{Aberle}},
  \bibinfo{author}{\bibfnamefont{S.}~\bibnamefont{Glunz}}, \bibnamefont{and}
  \bibinfo{author}{\bibfnamefont{W.}~\bibnamefont{Warta}},
  \bibinfo{journal}{Journal of Applied Physics} \textbf{\bibinfo{volume}{71}},
  \bibinfo{pages}{4422} (\bibinfo{year}{1992}).

\bibitem[{\citenamefont{Moison et~al.}(1986)\citenamefont{Moison, Van~Rompay,
  and Bensoussan}}]{moison1986}
\bibinfo{author}{\bibfnamefont{J.~M.} \bibnamefont{Moison}},
  \bibinfo{author}{\bibfnamefont{M.}~\bibnamefont{Van~Rompay}},
  \bibnamefont{and}
  \bibinfo{author}{\bibfnamefont{M.}~\bibnamefont{Bensoussan}},
  \bibinfo{journal}{Applied Physics Letters} \textbf{\bibinfo{volume}{48}},
  \bibinfo{pages}{1362} (\bibinfo{year}{1986}).

\bibitem[{\citenamefont{Aleksiejunas et~al.}(2003)\citenamefont{Aleksiejunas,
  Sudzius, Malinauskas, Vaitkus, Jarasiunas, and Sakai}}]{aleksiejunas2003}
\bibinfo{author}{\bibfnamefont{R.}~\bibnamefont{Aleksiejunas}},
  \bibinfo{author}{\bibfnamefont{M.}~\bibnamefont{Sudzius}},
  \bibinfo{author}{\bibfnamefont{T.}~\bibnamefont{Malinauskas}},
  \bibinfo{author}{\bibfnamefont{J.}~\bibnamefont{Vaitkus}},
  \bibinfo{author}{\bibfnamefont{K.}~\bibnamefont{Jarasiunas}},
  \bibnamefont{and} \bibinfo{author}{\bibfnamefont{S.}~\bibnamefont{Sakai}},
  \bibinfo{journal}{Applied Physics Letters} \textbf{\bibinfo{volume}{83}},
  \bibinfo{pages}{1157} (\bibinfo{year}{2003}).

\bibitem[{\citenamefont{Cusc{\'o} et~al.}(2009)\citenamefont{Cusc{\'o},
  Ib{\'a}{\~n}ez, Alarc{\'o}n-Llad{\'o}, Art{\'u}s, Yamaguchi, and
  Nanishi}}]{cusco2009}
\bibinfo{author}{\bibfnamefont{R.}~\bibnamefont{Cusc{\'o}}},
  \bibinfo{author}{\bibfnamefont{J.}~\bibnamefont{Ib{\'a}{\~n}ez}},
  \bibinfo{author}{\bibfnamefont{E.}~\bibnamefont{Alarc{\'o}n-Llad{\'o}}},
  \bibinfo{author}{\bibfnamefont{L.}~\bibnamefont{Art{\'u}s}},
  \bibinfo{author}{\bibfnamefont{T.}~\bibnamefont{Yamaguchi}},
  \bibnamefont{and} \bibinfo{author}{\bibfnamefont{Y.}~\bibnamefont{Nanishi}},
  \bibinfo{journal}{Physical Review B} \textbf{\bibinfo{volume}{80}},
  \bibinfo{pages}{155204} (\bibinfo{year}{2009}).

\bibitem[{\citenamefont{Wang et~al.}(1998)\citenamefont{Wang, Wong, Foreman,
  Yang, and Wong}}]{wang1998}
\bibinfo{author}{\bibfnamefont{H.}~\bibnamefont{Wang}},
  \bibinfo{author}{\bibfnamefont{K.}~\bibnamefont{Wong}},
  \bibinfo{author}{\bibfnamefont{B.}~\bibnamefont{Foreman}},
  \bibinfo{author}{\bibfnamefont{Z.}~\bibnamefont{Yang}}, \bibnamefont{and}
  \bibinfo{author}{\bibfnamefont{G.}~\bibnamefont{Wong}},
  \bibinfo{journal}{Journal of Applied Physics} \textbf{\bibinfo{volume}{83}},
  \bibinfo{pages}{4773} (\bibinfo{year}{1998}).

\bibitem[{\citenamefont{Cui et~al.}(2002)\citenamefont{Cui, Groza, Hillman,
  Burger, and James}}]{cui2002}
\bibinfo{author}{\bibfnamefont{Y.}~\bibnamefont{Cui}},
  \bibinfo{author}{\bibfnamefont{M.}~\bibnamefont{Groza}},
  \bibinfo{author}{\bibfnamefont{D.}~\bibnamefont{Hillman}},
  \bibinfo{author}{\bibfnamefont{A.}~\bibnamefont{Burger}}, \bibnamefont{and}
  \bibinfo{author}{\bibfnamefont{R.}~\bibnamefont{James}},
  \bibinfo{journal}{Journal of Applied Physics} \textbf{\bibinfo{volume}{92}},
  \bibinfo{pages}{2556} (\bibinfo{year}{2002}).

\bibitem[{\citenamefont{Aspnes}(1983)}]{aspnes1983}
\bibinfo{author}{\bibfnamefont{D.}~\bibnamefont{Aspnes}},
  \bibinfo{journal}{Surface Science} \textbf{\bibinfo{volume}{132}},
  \bibinfo{pages}{406} (\bibinfo{year}{1983}).

\bibitem[{\citenamefont{Saitoh and Hasegawa}(1992)}]{saitoh1992}
\bibinfo{author}{\bibfnamefont{T.}~\bibnamefont{Saitoh}} \bibnamefont{and}
  \bibinfo{author}{\bibfnamefont{H.}~\bibnamefont{Hasegawa}},
  \bibinfo{journal}{Applied Surface Science} \textbf{\bibinfo{volume}{56}},
  \bibinfo{pages}{94} (\bibinfo{year}{1992}).

\bibitem[{\citenamefont{Favorskiy et~al.}(2010)\citenamefont{Favorskiy, Vu,
  Peytavit, Arscott, Paget, and Rowe}}]{favorskiy2010}
\bibinfo{author}{\bibfnamefont{I.}~\bibnamefont{Favorskiy}},
  \bibinfo{author}{\bibfnamefont{D.}~\bibnamefont{Vu}},
  \bibinfo{author}{\bibfnamefont{E.}~\bibnamefont{Peytavit}},
  \bibinfo{author}{\bibfnamefont{S.}~\bibnamefont{Arscott}},
  \bibinfo{author}{\bibfnamefont{D.}~\bibnamefont{Paget}}, \bibnamefont{and}
  \bibinfo{author}{\bibfnamefont{A.}~\bibnamefont{Rowe}},
  \bibinfo{journal}{Rev. Sci. Instr.} \textbf{\bibinfo{volume}{81}},
  \bibinfo{pages}{103902} (\bibinfo{year}{2010}).

\bibitem[{\citenamefont{Vu et~al.}(2010)\citenamefont{Vu, Arscott, Peytavit,
  Ramdani, Gil, Andr{\'e}, Bansropun, G{\'e}rard, Rowe, and Paget}}]{vu2010}
\bibinfo{author}{\bibfnamefont{D.}~\bibnamefont{Vu}},
  \bibinfo{author}{\bibfnamefont{S.}~\bibnamefont{Arscott}},
  \bibinfo{author}{\bibfnamefont{E.}~\bibnamefont{Peytavit}},
  \bibinfo{author}{\bibfnamefont{R.}~\bibnamefont{Ramdani}},
  \bibinfo{author}{\bibfnamefont{E.}~\bibnamefont{Gil}},
  \bibinfo{author}{\bibfnamefont{Y.}~\bibnamefont{Andr{\'e}}},
  \bibinfo{author}{\bibfnamefont{S.}~\bibnamefont{Bansropun}},
  \bibinfo{author}{\bibfnamefont{B.}~\bibnamefont{G{\'e}rard}},
  \bibinfo{author}{\bibfnamefont{A.}~\bibnamefont{Rowe}}, \bibnamefont{and}
  \bibinfo{author}{\bibfnamefont{D.}~\bibnamefont{Paget}},
  \bibinfo{journal}{Physical Review B} \textbf{\bibinfo{volume}{82}},
  \bibinfo{pages}{115331} (\bibinfo{year}{2010}).

\bibitem[{\citenamefont{Kronik and Shapira}(1999)}]{kronik1999}
\bibinfo{author}{\bibfnamefont{L.}~\bibnamefont{Kronik}} \bibnamefont{and}
  \bibinfo{author}{\bibfnamefont{Y.}~\bibnamefont{Shapira}},
  \bibinfo{journal}{Surface Science Reports} \textbf{\bibinfo{volume}{37}},
  \bibinfo{pages}{1} (\bibinfo{year}{1999}).

\bibitem[{\citenamefont{Paget et~al.}(2012)\citenamefont{Paget, Cadiz, Rowe,
  Moreau, Arscott, and Peytavit}}]{paget2012}
\bibinfo{author}{\bibfnamefont{D.}~\bibnamefont{Paget}},
  \bibinfo{author}{\bibfnamefont{F.}~\bibnamefont{Cadiz}},
  \bibinfo{author}{\bibfnamefont{A.}~\bibnamefont{Rowe}},
  \bibinfo{author}{\bibfnamefont{F.}~\bibnamefont{Moreau}},
  \bibinfo{author}{\bibfnamefont{S.}~\bibnamefont{Arscott}}, \bibnamefont{and}
  \bibinfo{author}{\bibfnamefont{E.}~\bibnamefont{Peytavit}},
  \bibinfo{journal}{Journal of Applied Physics} \textbf{\bibinfo{volume}{111}},
  \bibinfo{pages}{123720} (\bibinfo{year}{2012}).

\bibitem[{\citenamefont{Stephens et~al.}(1994)\citenamefont{Stephens, Aberle,
  and Green}}]{stephens1994}
\bibinfo{author}{\bibfnamefont{A.}~\bibnamefont{Stephens}},
  \bibinfo{author}{\bibfnamefont{A.}~\bibnamefont{Aberle}}, \bibnamefont{and}
  \bibinfo{author}{\bibfnamefont{M.}~\bibnamefont{Green}},
  \bibinfo{journal}{Journal of Applied Physics} \textbf{\bibinfo{volume}{76}},
  \bibinfo{pages}{363} (\bibinfo{year}{1994}).

\bibitem[{\citenamefont{Visvanathan and Battey}(1954)}]{visvanathan1954}
\bibinfo{author}{\bibfnamefont{S.}~\bibnamefont{Visvanathan}} \bibnamefont{and}
  \bibinfo{author}{\bibfnamefont{J.}~\bibnamefont{Battey}},
  \bibinfo{journal}{Journal of Applied Physics} \textbf{\bibinfo{volume}{25}},
  \bibinfo{pages}{99} (\bibinfo{year}{1954}).

\bibitem[{\citenamefont{Zerrouati et~al.}(1988)\citenamefont{Zerrouati, Fabre,
  Bacquet, Bandet, Frandon, Lampel, and Paget}}]{zerrouati1988}
\bibinfo{author}{\bibfnamefont{K.}~\bibnamefont{Zerrouati}},
  \bibinfo{author}{\bibfnamefont{F.}~\bibnamefont{Fabre}},
  \bibinfo{author}{\bibfnamefont{G.}~\bibnamefont{Bacquet}},
  \bibinfo{author}{\bibfnamefont{J.}~\bibnamefont{Bandet}},
  \bibinfo{author}{\bibfnamefont{J.}~\bibnamefont{Frandon}},
  \bibinfo{author}{\bibfnamefont{G.}~\bibnamefont{Lampel}}, \bibnamefont{and}
  \bibinfo{author}{\bibfnamefont{D.}~\bibnamefont{Paget}},
  \bibinfo{journal}{{Phys. Rev. B}} \textbf{\bibinfo{volume}{37}},
  \bibinfo{pages}{1334} (\bibinfo{year}{1988}).

\bibitem[{\citenamefont{Paget et~al.}(1996)\citenamefont{Paget, Gusev, and
  Berkovits}}]{paget1996}
\bibinfo{author}{\bibfnamefont{D.}~\bibnamefont{Paget}},
  \bibinfo{author}{\bibfnamefont{A.}~\bibnamefont{Gusev}}, \bibnamefont{and}
  \bibinfo{author}{\bibfnamefont{V.}~\bibnamefont{Berkovits}},
  \bibinfo{journal}{Physical Review B} \textbf{\bibinfo{volume}{53}},
  \bibinfo{pages}{4615} (\bibinfo{year}{1996}).

\bibitem[{\citenamefont{Kang et~al.}(1997)\citenamefont{Kang, Sa, Park, Suh,
  and Oh}}]{kang1997}
\bibinfo{author}{\bibfnamefont{M.}~\bibnamefont{Kang}},
  \bibinfo{author}{\bibfnamefont{S.}~\bibnamefont{Sa}},
  \bibinfo{author}{\bibfnamefont{H.}~\bibnamefont{Park}},
  \bibinfo{author}{\bibfnamefont{K.}~\bibnamefont{Suh}}, \bibnamefont{and}
  \bibinfo{author}{\bibfnamefont{K.}~\bibnamefont{Oh}}, \bibinfo{journal}{Thin
  Solid Films} \textbf{\bibinfo{volume}{308}}, \bibinfo{pages}{634}
  (\bibinfo{year}{1997}).

\bibitem[{\citenamefont{Leong et~al.}(1996)\citenamefont{Leong, McMurray,
  Williams, and Stringfellow}}]{leong1996}
\bibinfo{author}{\bibfnamefont{J.}~\bibnamefont{Leong}},
  \bibinfo{author}{\bibfnamefont{J.}~\bibnamefont{McMurray}},
  \bibinfo{author}{\bibfnamefont{C.}~\bibnamefont{Williams}}, \bibnamefont{and}
  \bibinfo{author}{\bibfnamefont{G.}~\bibnamefont{Stringfellow}},
  \bibinfo{journal}{Journal of Vacuum Science \& Technology B: Microelectronics
  and Nanometer Structures} \textbf{\bibinfo{volume}{14}},
  \bibinfo{pages}{3113} (\bibinfo{year}{1996}).

\bibitem[{\citenamefont{Sasaki et~al.}(2001)\citenamefont{Sasaki, Tsuchida,
  Narukawa, Kawakami, Fujita, Hsu, and Stringfellow}}]{sasaki2001}
\bibinfo{author}{\bibfnamefont{A.}~\bibnamefont{Sasaki}},
  \bibinfo{author}{\bibfnamefont{K.}~\bibnamefont{Tsuchida}},
  \bibinfo{author}{\bibfnamefont{Y.}~\bibnamefont{Narukawa}},
  \bibinfo{author}{\bibfnamefont{Y.}~\bibnamefont{Kawakami}},
  \bibinfo{author}{\bibfnamefont{S.}~\bibnamefont{Fujita}},
  \bibinfo{author}{\bibfnamefont{Y.}~\bibnamefont{Hsu}}, \bibnamefont{and}
  \bibinfo{author}{\bibfnamefont{G.}~\bibnamefont{Stringfellow}},
  \bibinfo{journal}{Journal of Applied Physics} \textbf{\bibinfo{volume}{89}},
  \bibinfo{pages}{343} (\bibinfo{year}{2001}).

\bibitem[{\citenamefont{Rhoderick}(1982)}]{rhoderick1982}
\bibinfo{author}{\bibfnamefont{E.}~\bibnamefont{Rhoderick}},
  \emph{\bibinfo{title}{Metal semiconductor contacts}}
  (\bibinfo{publisher}{IET}, \bibinfo{year}{1982}).

\bibitem[{\citenamefont{Henry and Lang}(1977)}]{henry1977}
\bibinfo{author}{\bibfnamefont{C.}~\bibnamefont{Henry}} \bibnamefont{and}
  \bibinfo{author}{\bibfnamefont{D.}~\bibnamefont{Lang}},
  \bibinfo{journal}{Physical Review B} \textbf{\bibinfo{volume}{15}},
  \bibinfo{pages}{989} (\bibinfo{year}{1977}).

\bibitem[{\citenamefont{Deenapanray et~al.}(2003)\citenamefont{Deenapanray,
  Tan, and Jagadish}}]{deenapanray2003}
\bibinfo{author}{\bibfnamefont{P.}~\bibnamefont{Deenapanray}},
  \bibinfo{author}{\bibfnamefont{H.}~\bibnamefont{Tan}}, \bibnamefont{and}
  \bibinfo{author}{\bibfnamefont{C.}~\bibnamefont{Jagadish}},
  \bibinfo{journal}{Applied Physics A: Materials Science \& Processing}
  \textbf{\bibinfo{volume}{76}}, \bibinfo{pages}{961} (\bibinfo{year}{2003}).

\bibitem[{\citenamefont{Brunkov et~al.}(2000)\citenamefont{Brunkov, Chaldyshev,
  Chernigovskii, Suvorova, Bert, Konnikov, Preobrazhenskii, Puyato, and
  Semyagin}}]{brunkov2000}
\bibinfo{author}{\bibfnamefont{P.}~\bibnamefont{Brunkov}},
  \bibinfo{author}{\bibfnamefont{V.}~\bibnamefont{Chaldyshev}},
  \bibinfo{author}{\bibfnamefont{A.}~\bibnamefont{Chernigovskii}},
  \bibinfo{author}{\bibfnamefont{A.}~\bibnamefont{Suvorova}},
  \bibinfo{author}{\bibfnamefont{N.}~\bibnamefont{Bert}},
  \bibinfo{author}{\bibfnamefont{S.}~\bibnamefont{Konnikov}},
  \bibinfo{author}{\bibfnamefont{V.}~\bibnamefont{Preobrazhenskii}},
  \bibinfo{author}{\bibfnamefont{M.}~\bibnamefont{Puyato}}, \bibnamefont{and}
  \bibinfo{author}{\bibfnamefont{B.}~\bibnamefont{Semyagin}},
  \bibinfo{journal}{Semiconductors} \textbf{\bibinfo{volume}{34}},
  \bibinfo{pages}{1068} (\bibinfo{year}{2000}).

\end{thebibliography}

\end{document}